# Leveraging GANs for data scarcity of COVID-19: Beyond the hype

Hazrat Ali[1], Christer Grönlund[2], Zubair Shah[1]

[1] College of Science and Engineering, Hamad Bin Khalifa University, Qatar Foundation, Doha, Qatar

[2] Department of Radiation Sciences, Umea University, Umea, Sweden.

## Abstract

Artificial Intelligence (AI)-based models can help in diagnosing COVID-19 from lung CT scans and X-ray images; however, these models require large amounts of data for training and validation. Many researchers studied Generative Adversarial Networks (GANs) for producing synthetic lung CT scans and X-Ray images to improve the performance of AI-based models. It is not well explored how good GAN-based methods performed to generate reliable synthetic data. This work analyzes 43 published studies that reported GANs for synthetic data generation. Many of these studies suffered data bias, lack of reproducibility, and lack of feedback from the radiologists or other domain experts. A common issue in these studies is the unavailability of the source code, hindering reproducibility. The included studies reported rescaling of the input images to train the existing GANs architecture without providing clinical insights on how the rescaling was motivated. Finally, even though GAN-based methods have the potential for data augmentation and improving the training of AI-based models, these methods fall short in terms of their use in clinical practice. This paper highlights research hotspots in countering the data scarcity problem, identifies various issues as well as potentials, and provides recommendations to guide future research. These recommendations might be useful to improve acceptability for the GAN-based approaches for data augmentation as GANs for data augmentation are increasingly becoming popular in the AI and medical imaging research community.

## Introduction

The healthcare systems worldwide faced an unprecedented challenge with the COVID-19 pandemic. As a result, the capacity of coping with faster testing and providing care was put to test. When COVID-19 spread wildly, researchers were pushed to find quick ways for developing AI techniques to aid in combating the pandemic through early diagnosis.

The most promising Artificial Intelligence (AI) techniques mostly fall in the deep learning category – models that consists of multiple layers of neural network (for example, convolutional neural networks (CNNs) and their variants). However, the potential of deep learning models to learn from the data rely on very large data. For image-based diagnosis and analysis, these models required a large amount of lungs radiology images data. Since the data availability was meager, researchers rushed to use Generative Adversarial Networks (GANs) to generate synthetic Computed Tomography (CT) scans or X-Ray images that may capture the characteristics of real data with COVID-19 signs. It is well understood that image data augmentation is the most common application where GANs have found promising use due to their ability to generate realistic-looking images. Consequently, many studies reported the use of GANs to combat the data scarcity problem in training AI models for COVID-19 diagnosis (for example, this review identified 43 studies [1 – 43]. Typically, GANs serve as a sub-module of the entire framework used explicitly as a data augmentation method, while the diagnosis of COVID-19 is made by using appropriate AI methods, for example, convolutional neural networks (ResNet, VGG16, etc) [44], [45].

**Problem statement**: While many reviews have been published on the role of AI methods in COVID-19 diagnosis [44 – 47], these reviews did not explicitly cover the shortcomings and risk of bias of the studies or did not explicitly cover GANs methods. One particularly relevant work is the review by Robert et al. [48] however, that has a broad scope as it covers various machine learning methods for COVID-19. Besides, the focus in [48] is mainly the performance of AI methods for the diagnosis of COVID-19 and does not identify the limitations of the data augmentation methods explicitly. Another relevant study is the review by [49] however, the focus of their work was on reviewing X-ray datasets for COVID-19 and not on reviewing the models. With the growing number of studies on GANs for synthetic data for COVID-19, it is important to review the reported methods and analyze them, particularly from the perspectives of model generalization, data representation, and clinical translation. Furthermore, to enhance the research and developments on GANs based methods, it is critical to understand the importance of the risk of data bias, the associated challenges to the diagnosis, the evaluation mechanisms, and the inclusion of radiologists or domain experts in the loop when the research community presents more developments on the topic.

This review aims to appreciate the early attempts for GANs-based methods to address the challenges related to COVID-19 data scarcity and diagnosis. Besides, it seeks to highlight the importance of addressing specific blights and imperfections as overcoming these can help increase the effectiveness and usability of the findings of these research developments. Unlike previous reviews, that cover a general description of the strengths of the AI methods for COVID-19 diagnosis, this work provides a more in-depth description of the various challenges that are associated with the GANs-based augmentation of data for COVID-19.

**Methods and Materials:**

We performed a literature search from 12 October 2021 to 13 October 2021 to retrieve relevant published studies that reported the use of GANs for the generating lung CT scan or X-Ray image data. We retrieved the studies from Pubmed, Scopus, IEEEXplore, and Google Scholar. We found a total of 348 studies. In the first phase of study selection, we removed 81 duplicates and then performed title and abstract screening and removed 208 studies. We performed the full-text reading in the second phase and excluded 16 studies that did not fulfill the inclusion criteria (as identified in supplementary material). Finally, we included 43 studies in this analysis. We cover only those studies that reported GANs for synthesis (data augmentation) purposes. Appendix 1 provides the search terms used for retrieval of the studies. Appendix 2 provides the criteria for inclusion and exclusion of the studies.

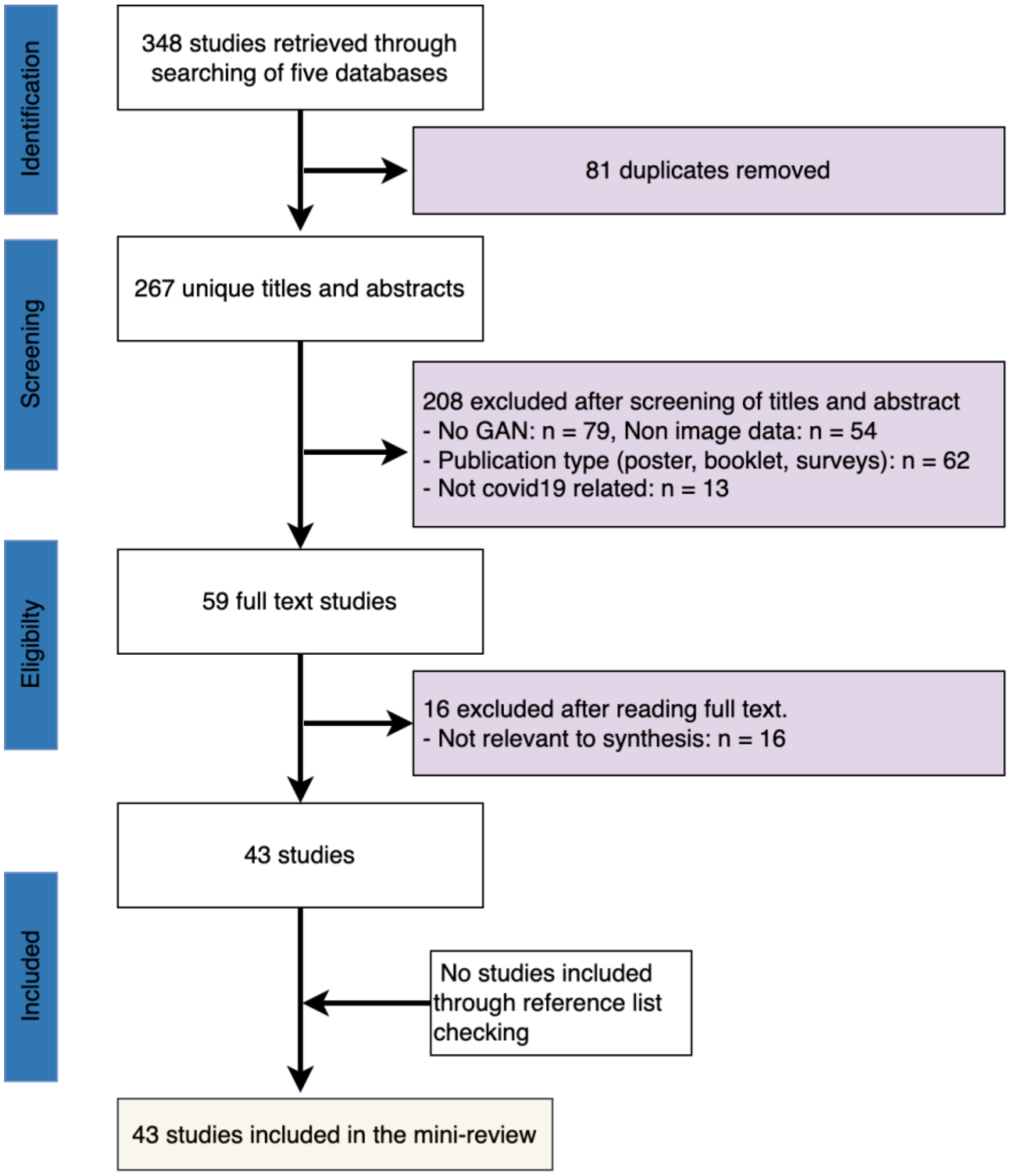

Figure 1. PRISMA flowchart for the selection of studies.

**Results and Discussion:**

Figure 1 shows the PRISMA flowchart for the selection of studies and Figure 2 shows a summary of the demographics of the included studies. The key findings of our analysis can be categorized into the challenges related to:

1. Data proportion, such as the dataset size used for model training, the underlying bias in the data or the model training, and the associated data leakage problems.
2. The quality of the data such as the image resolution and the data modality.
3. The applications in COVID-19 such as the lack of demonstration of diagnosis.

4. The evaluation mechanism such as qualitative evaluation by radiologists or the metrics used for quantitative evaluation.
5. The potential clinical translations such as code availability and reproducibility.

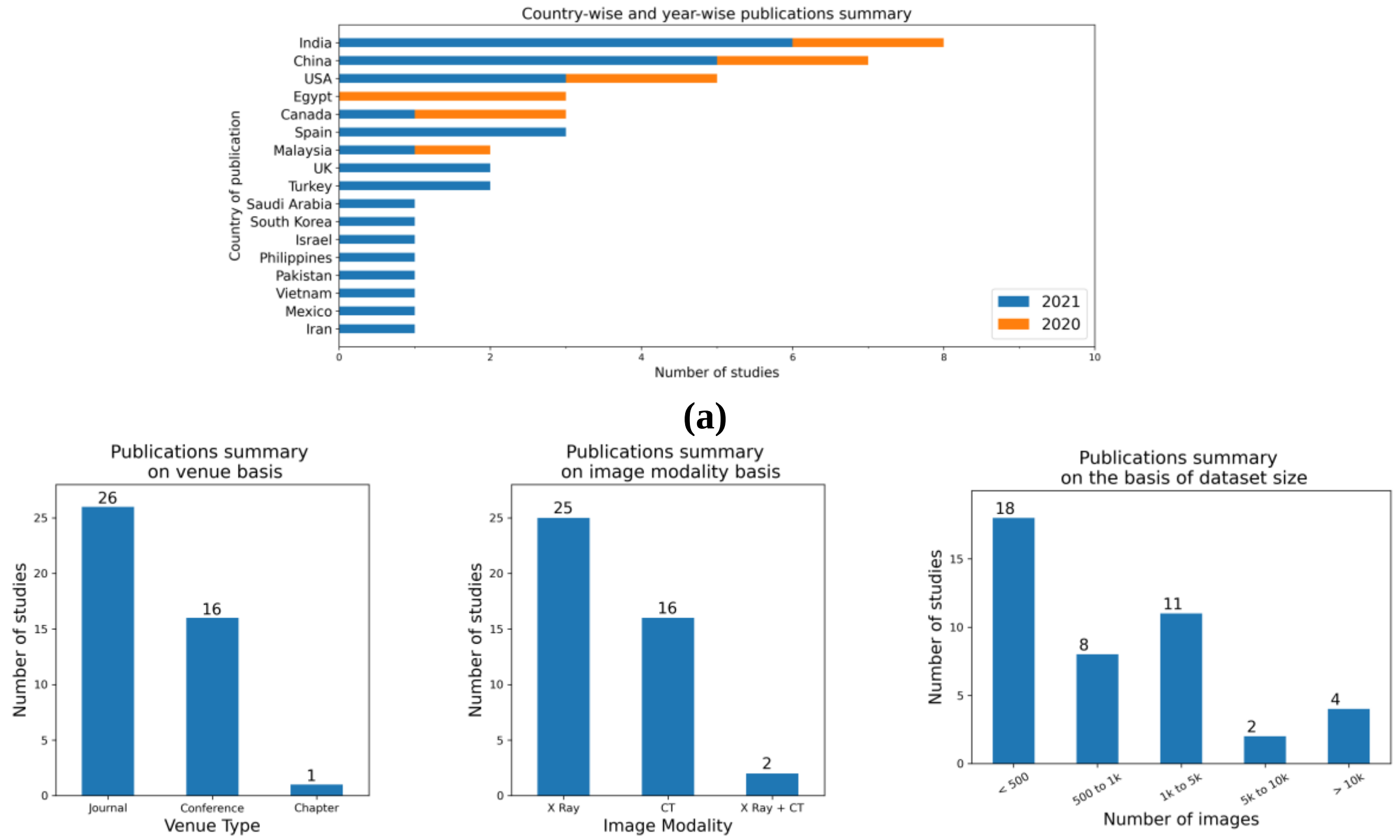

**(b)** **(c)** **(d)**

Figure 2. Demographics of the studies. (a) Summary of country-wise and year-wise number of studies. (b) Summary of number of the studies on the basis of venue. (c) Summary of number of studies on the basis of using different modalities of training data. (d) Summary of studies on the basis of the number of images used for training.

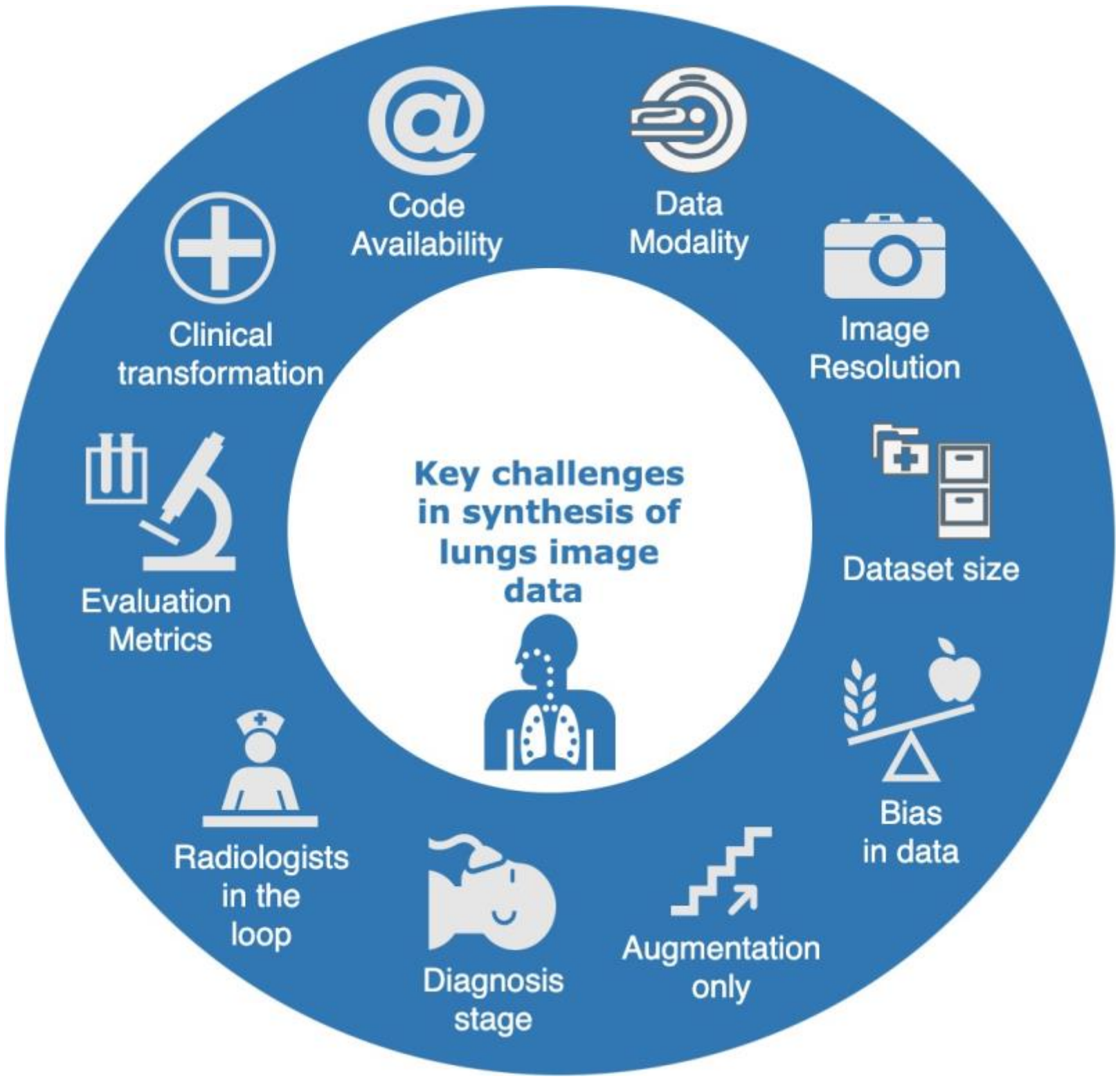

Figure 3. Key challenges identified in the studies reporting GANs based data augmentation for COVID-19

Figure 3 provides a summary of the challenges identified in this analysis. We discuss these in the following text.

**Dataset size**: Better representation of the data can be obtained from large data. However, in the included studies, only four studies [17], [36], [39], [43] reported using more than ten thousand images, and only two studies [12], [37] reported using more than 5000 images. In more than half of the studies [1-7], [14-16], [20-29], [31], [34], [35], [38], [40], [42], the number of images reported for training was less than 1000. So, this training data size raises concerns about the generalization of GANs training. Similarly, in many studies, we found a lack of information on the number of patients (individuals) from whom the images have been acquired. Besides, we found that the number of individuals, even if reported in the studies, was too small compared to what would be recommended for training a GAN model. Only three studies [4], [7], [20] reported using data of more than 500 individuals. In addition, a common challenge in the crowd-source data is mapping the number of images to the number of individuals. Furthermore, the inclusion/exclusion of an individual and the methods and criteria for recruiting individuals to record data may affect the data demographics, bias, and imbalance, consequently influencing the model training.

**Risk of bias**: Bias in data refers to the imbalanced representation of different groups in the data. For example, there can be gender bias, age bias, or sample bias, etc. If the original data used for training GANs has a bias, then there is a high risk that the bias will be carried to the synthetic data. It is clear that the crowd-sourced datasets available publicly lacked any such specifications, or the studies fail to report to cater for the biases in the data. Eventually, the bias in data will result in a bias in the training of the AI-based diagnosis model too. For example, Garcia et al. [49] reported in their analysis of studies reporting X-ray images datasets to diagnose COVID-19 that the models trained with chest X-ray datasets were prone to high bias.

The potential risk of bias in training AI models is a critical bottleneck in the generalization ability and robustness of the models. For example, at the beginning of the pandemic, most of the positive cases were for adults only. On the other hand, a pneumonia dataset may have more samples for lung X-ray images of children. A GANs model trained on such data will carry forward a similar bias. Suppose we use the pneumonia dataset with more young people/pediatrics samples and the COVID-19 dataset with more examples for adults. In that case, the training may end up in a model that has learned to classify lung images of adults versus children instead of classifying COVID-19 versus pneumonia.

Some studies [4], [22], [29], [32] used the RSNA pneumonia dataset available from Kaggle[1]. This particular dataset, as reported by Kermany et al., [50] has image data for young people between the age of one to five. Ablation studies and adversarial tests are important to report to avoid such biases. So, one may test the model for extreme cases such as samples of adults with pneumonia or samples of young people that were COVID-19 positive.

**Data leakage**: Some studies report results for multiple datasets. However, no cross-verification mechanism was reported to avoid data leakage. Hence, reporting results on two different public datasets does not necessarily imply validation/testing on independent samples as these datasets may have borrowed from each other. For example, [11], [18], [33] reported the use of three different datasets of chest X-ray images. However, the data in Tawsif et al., [44] borrows images from Cohen et al., [51]. Amongst these, [18] reported deletion of duplicate images though details were missing on how duplicate images were identified). [11] reported that they used one dataset to train the model and then used the trained model to label images of the other dataset but did not identify that the two datasets have overlapping entries. In summary, prevention of data leakage is not guaranteed by the studies, and the studies lack explicit details on addressing the confounders. The majority of these studies (36 out of 43 studies) reported using a publicly available dataset for CT and X-ray images. For similar reasons as discussed above, many of the images in these data were collected through crowdsourcing as an early attempt to facilitate research on AI for COVID-19. Consequently, the data collection (more often through crowdsourcing) was mostly uncontrolled and not well-curated. Hence, they lacked many details such as demographics or guarantee to prevent data leakage.

---

[1] https://www.kaggle.com/c/rsna-pneumonia-detection-challenge/data

Only a few studies [7], [12], [14], [15], [20], [31], [34] reported using privately collected data where the researchers had more control over the data collection mechanism.

**Image resolution**: A common matter found in the included studies was rescaling the image data to a resolution 512 × 512 or 224 × 224 pixels. The motivation for resizing the images was mostly model-driven and not driven by the clinical objective. At least, the studies provided no insights on how or why they chose this resolution. In some cases, the rescaling seems quite an aggressive downsizing of the input X-ray images from 2170 × 1953 pixels to 512 × 512. Similarly, for volumetric CT scans, the studies had to restrict themselves to 2D images (individual scans) to train the model. Insights on such choices are either missing or very limited.

**Data modality:** In the 43 studies, we found that almost 60% of the studies (25 used X-ray images versus 16 used CT) reported using X-ray image data to train GANs models for data augmentation to achieve diagnosis with AI models. However, it is well understood that CT scan have higher sensitivity than X-ray images for COVID-19 diagnosis [49].

**Augmentation only**: Some studies reported using GAN-based methods to generate synthetic lung CT scans or X-ray images [1], [14], [21], [22], [24], [25], [34], [35], [42]. For example, [14] covers segmentation only). [25] reports the synthesis of volumetric CT scans using a 3D conditional GAN however, the study did not analyze the generated synthetic data for its ability for diagnosing COVID-19, meaning the studies did not explore the effectiveness of the generated synthetic data.

**Diagnosis stage:** One of the common trends in studies was the staging of the disease. Most of the image data shared during the earlier days of the pandemic comprised cases of severe COVID-19 with a noticeable impact on the lungs. Consequently, the GANs trained would generate synthetic images of the extreme stage where the presence of the infection was already established. Hence, GANs-based methods that can support early diagnosis of the infection are limited.

**Radiologists in the loop**: In the included studies, none (but one study [22]) reported an evaluation/rating of the synthesized data by presenting it to radiologists. So, there remains an unclear interpretation of the qualitative assessment of the synthesized CT/X-Ray images. [22] reported qualitative analysis of synthesized X-ray images by a radiologist that the radiologist could distinguish from real X-ray images for COVID-19 positive cases. Besides, the qualitative analysis suggested that the synthesized images fall short of diagnosis quality.

**Evaluation Metrics:** One common mechanism for quantitative evaluation of synthetic CT scan or X-ray images was the use of metrics such as Structure Similarity Index Measure (SSIM), Frechet Inception Distance (FID), and Peak Signal-to-Noise Ratio (PSNR). However, the evaluation with SSIM and FID without input from radiologists can be overly optimistic as these metrics are primarily derived from computer vision literature, and their suitability for medical images might be limited. Besides, the lack of evaluation by radiologists might hinder the acceptability of these models being translated into clinical practice and hence lose the very purpose of using AI to aid in combating the pandemic.

**Code availability**: To advance the developments in data augmentation, there is a dire need to provide reproducible softwares or code. We found the lack of reproducibility analysis as a common trend in the studies. More specifically, only three studies [14], [24], [39] provided links to publicly available Github repositories for their code. Among these, one of the links [14] was found broken.

**Can the studies be translated to clinical applications**: We believe with the existing shortcomings in the studies, GANs based studies are not yet ready to be translated to clinical practice. Nevertheless, despite the shortcomings, the potential impact of using GAN-based methods to improve the training of AI models for COVID-19 diagnosis cannot be denied. Therefore, future transformation to clinical applications is not futile.

**Suggestions**

In this section, we provide suggestions on handling the GAN-based data augmentation for COVID-19. We believe that these suggestions will provide a roadmap to the research community and improve future study designs.

**Large data:** Since much more data is becoming available for lungs CT scan and X-Ray images; it is highly recommended that any current and future research on GANs for COVID-19 is based on much larger datasets than those reported in the earlier studies surveyed in this work. In addition, the reporting of data demographics should be encouraged as this will help mitigate challenges such as data leakage or risk of bias in data.

**Reproducibility:** We urge the research community to publish their codes for reproducibility to promote usability and utility. The transparency in using the models will be a key to increasing their acceptability and help the community gain better insights into the model. In addition, it will also enable reproducibility as well as future developments on top of existing works.

**Adversarial test:** A useful way to evaluate the model is to report aggressive adversarial test strategies, for example, by generating GANs based data from lungs CT scans before 2020 and then using the generating data to evaluate the AI model for COVID-19 diagnosis. Any positive detection will provide an opportunity to investigate the model performance as COVID-19 was only spreading in late 2020.

**Evaluation methods:** The research community (including the authors, the reviewers, and the readers) should consider evaluating their work against the Radiological Society of North America Checklist (RSNA CLAIM) [52], assessment tools like PROBAST [53], or similar metrics, that provide a comprehensive checklist for AI models from the perspective of data, model training, and evaluation metrics. A lack of compliance with the guidelines such as those in CLAIM [52] or validation of the studies using assessment tools such as PROBAST [53], will hamper the translation of these findings into clinical applications.

**Early feedback from radiologists:** The research community should consider incorporating the feedback and input from doctors and clinicians starting from the earlier phase of the study design. Doing so will help better understand the key challenges in handling lung CT scans or X-ray images data, interpret the subtle information in the data, and increase the opportunities for clinical transformation of the developed methods.

**Limitations**: Our analysis includes studies from four databases, namely Pubmed, Scopus, IEEE Xplore, and Google Scholar. So, studies that are not indexed in these databases have been left out. We did not include pre-prints and unpublished literature in this review as they are not peer-reviewed, but might contain good research outcomes. The scope of our review is limited to images-based studies only. The analysis does not directly compare the evaluation metrics as the included studies differ in terms of the dataset size, choice of GAN architecture, the timeline of the study. The research and developments on methods for COVID-19 are extremely fast. Hence, it is possible that as this mini-review is being written, several additional studies might be published which are not covered in this analysis.

**Conclusion**: This mini-review provided a critical analysis of the shortcomings of GANs-based methods for data augmentation in applications of diagnosing COVID-19. We identified many areas where the research on using GANs-based methods for data augmentation in COVID-19 could be improved. We believe that the findings of this analysis of studies reporting GANs-based synthetic data complement findings of existing review by [48] who reported the many pitfalls in the use of machine learning methods for COVID-19, and an earlier study by [49] that identified the risk of bias in the reported AI models. The analysis in this mini-review will help the readers to understand the limitations of published studies and design better studies in the future to overcome the shortcomings.

**Author Contributions:**

H.A contributed to the conception, design, literature search, data selection, data synthesis, data extraction, drafting. C.G contributed to the design and critical revision of the manuscript. Z.S contributed to the conception, design, and critical revision of the manuscript. All authors gave their final approval and accepted accountability for all aspects of the work.

**Competing Interests:**

The authors declare no competing financial or non-financial interests.